Progesterone: An Enigmatic Ligand for the Mineralocorticoid Receptor

Michael E. Baker[1]
Yoshinao Katsu[2]

[1]Division of Nephrology-Hypertension
Department of Medicine, 0735
University of California, San Diego
9500 Gilman Drive
La Jolla, CA 92093-0735
[2]Graduate School of Life Science
Hokkaido University
Sapporo, Japan

Correspondence to:
M. E. Baker; E-mail: mbaker@ucsd.edu

Y. Katsu; E-mail: ykatsu@sci.hokudai.ac.jp

**Abstract.** The progesterone receptor (PR) mediates progesterone regulation of female reproductive physiology, as well as gene transcription in non-reproductive tissues, such as brain, bone, lung and vasculature, in both women and men. An unusual property of progesterone is its high affinity for the mineralocorticoid receptor (MR), which regulates electrolyte transport in the kidney in humans and other terrestrial vertebrates. In humans, rats, alligators and frogs, progesterone antagonizes activation of the MR by aldosterone, the physiological mineralocorticoid in terrestrial vertebrates. In contrast, in elephant shark, ray-finned fishes and chickens, progesterone activates the MR. Interestingly, cartilaginous fishes and ray-finned fishes do not synthesize aldosterone, raising the question of which steroid(s) activate the MR in cartilaginous fishes and ray-finned fishes. The simpler synthesis of progesterone, compared to cortisol and other corticosteroids, makes progesterone a candidate physiological activator of the MR in elephant sharks and ray-finned fishes. Elephant shark and ray-finned fish MRs are expressed in diverse tissues, including heart, brain and lung, as well as, ovary and testis, two reproductive tissues that are targets for progesterone, which together suggests a multi-faceted physiological role for progesterone activation of the MR in elephant shark and ray-finned fish. The functional consequences of progesterone as an antagonist of some terrestrial vertebrate MRs and as an agonist of fish and chicken MRs are not fully understood. Indeed, little is known of physiological activities of progesterone via any vertebrate MR.

**Introduction.** The progesterone receptor (PR) and mineralocorticoid receptor (MR) are steroid receptors that belong to the nuclear receptor family, a diverse group of transcription factors that also includes the estrogen receptor (ER), androgen receptor (AR), and glucocorticoid receptor (GR) [1-5]. In humans and other terrestrial vertebrates, the PR and MR mediate the physiological actions of progesterone and aldosterone, respectively (Figure 1). Although aldosterone is the main physiological mineralocorticoid in terrestrial vertebrates, diverse corticosteroids including cortisol, corticosterone and 11-deoxycorticosterone also are ligands for vertebrate MRs [6-8].

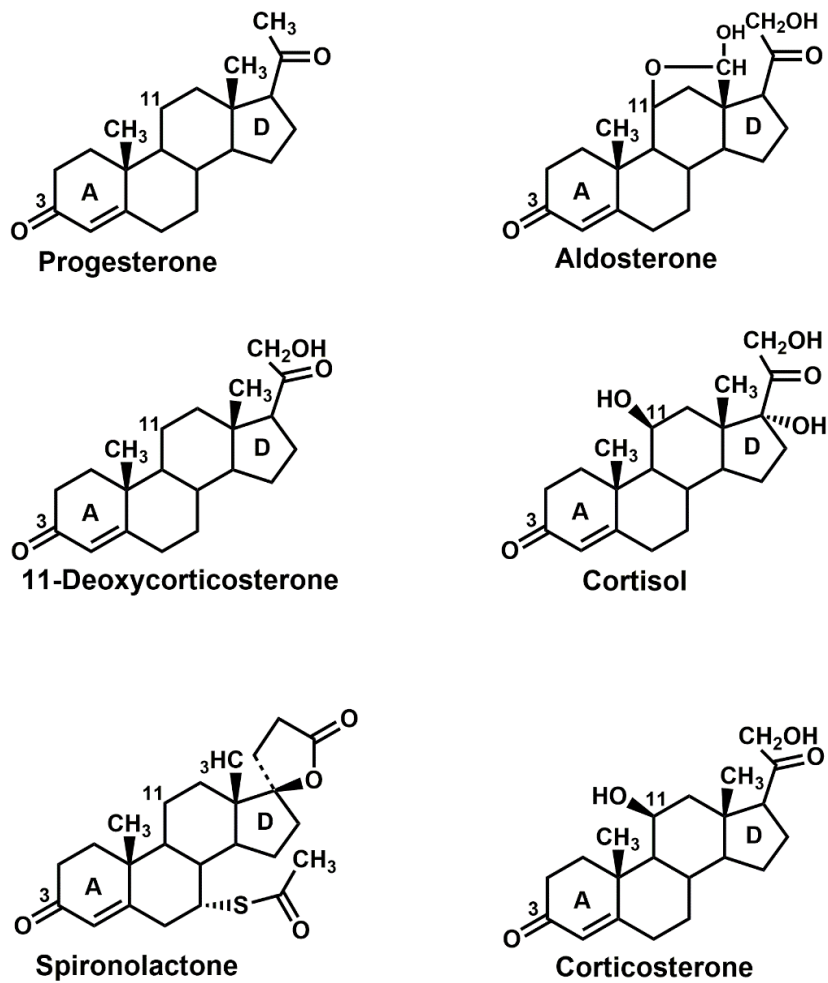

**Figure 1. Structures of Progesterone, Adrenal Steroids and Spironolactone.**
Aldosterone and 11-deoxycorticosterone are the main physiological mineralocorticoids in vertebrates. Cortisol and corticosterone are the main physiological glucocorticoids in vertebrates. Progesterone, a female reproductive steroid, and spironolactone are antagonists for human MR [9, 10] and agonists for fish MRs [11-16] and chicken MR [12, 13].

Progesterone and aldosterone have different functions in vertebrates. Progesterone is considered to be a female reproductive steroid due to its regulation of gene expression in mammary gland and uterus, which affects many aspects of female reproductive physiology, including fertilization, maintenance of pregnancy and preparation of the endometrium for implantation and parturition [17-20]. However, progesterone also has important physiological actions in males including in prostate and testes [21-24]. Moreover, progesterone activates the PR in the brain, bone, thymus, lung and vasculature in females and males [25, 26]. Thus, progesterone is a steroid with diverse physiological activities in many organs in females and males.

Aldosterone is the physiological mineralocorticoid in terrestrial vertebrates. Aldosterone activates the MR in the distal tubule of the kidney to retain sodium and regulate blood pressure and secret potassium [7, 27-32]. In addition to this classical function, aldosterone activation of the MR regulates electrolyte transport across other epithelial tissues, including colon, salivary and sweat glands and airway epithelia of the lung [29, 33-35]. Moreover, aldosterone activates the MR in nonepithelial tissues, including brain, heart, ovaries, pancreas, adipocytes [6, 33, 35-41]. Thus, like the PR, the MR is active in many organs with diverse physiological activities.

**Low concentrations of aldosterone and progesterone activate their receptors**

At concentrations below 1 nM, aldosterone and progesterone are transcriptional activators of the MR and PR, respectively. The half-maximal response (EC50) for transcriptional activation by aldosterone of human MR is about 0.1 nM [9, 13, 16, 42, 43]. EC50 values for transcriptional activation by progesterone of human PR are about 0.5 to 1 nM [44, 45]. Based on the high affinity of aldosterone and progesterone for the MR and PR, each steroid would be expected to be specific for its cognate receptor. Although this is true for aldosterone, which is specific for the MR, unexpectedly, as described below, progesterone has a high affinity for the MR.

**Progesterone: An unexpected high affinity antagonist for human MR**

Beginning in the 1950s, various *in vivo* studies found that, at nM concentrations, progesterone can antagonize aldosterone or 11-deoxycorticosterone activation of rat MR [46-48], human MR [49] and amphibian MR [50, 51]. These studies indicated that although progesterone had a high affinity for the MR, progesterone was not a transcriptional activator of the MR. Because *in vivo* studies of binding and activation of the MR by corticosteroids and progesterone can be complicated by their binding to serum proteins [8], purified MR was needed for an accurate determination of the affinity of progesterone for human MR. Purified MR became available with the cloning of the human MR by Arriza et al, [6]. Using recombinant human MR, Arriza et al reported that aldosterone, cortisol, corticosterone, 11-deoxycorticosterone, and progesterone had a similar nM affinity for human MR. Further studies by Rupprecht et al. [10], Geller et al. [9] and Katsu et al. [13] showed that progesterone was a human MR antagonist with sub nM affinity.

**Progesterone is a transcriptional activator of ray-finned fish and chicken MRs**

In contrast to the antagonist activity of progesterone for human, rodent and amphibian MRs, progesterone is a transcriptional activator of several fish MRs, including trout [15], zebrafish [11, 13, 52], sturgeon and gar [16]. Progesterone also activates chicken MR [12, 13]. Progesterone has EC50s of less than 1 nM for these MRs, which makes progesterone a potential physiological activator of these MRs.

**Progesterone is a transcriptional activator of elephant shark MR.**

The evidence that progesterone activates ray-finned fish MR, but not human MR raised the question of when did progesterone activation of ray-finned fish MRs evolve? Did progesterone activation of the MR evolve in the ray-finned fish line, or did it evolve in a cartilaginous fish, which belong to a group of vertebrates that diverged from ray-finned fish and terrestrial vertebrates about 450 million years ago [1, 36, 53-56] (Figure 2). To answer this question, we cloned elephant shark MR and studied its activation by progesterone, aldosterone and other corticosteroids [12]. We chose elephant shark MR because elephant shark occupies a key phylogenetic position in the evolution of ray-finned fish and terrestrial vertebrates [54, 56] (Figure 2). Moreover, genomic analyses reveal that elephant shark has the slowest evolving genome of all known vertebrates [54], including the coelacanth, making elephant shark MR an attractive receptor to study early events in the evolution of mechanisms for regulating MR transcription.

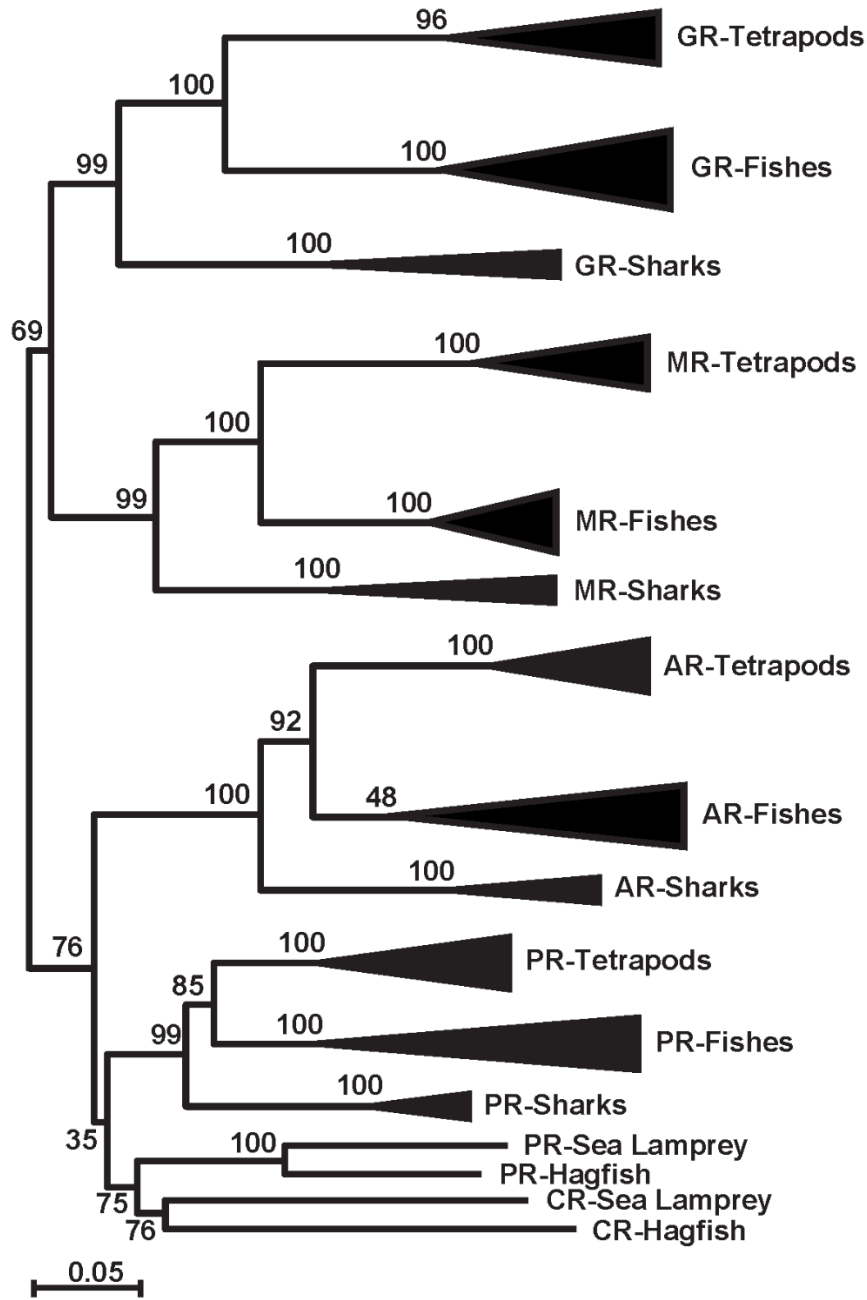

**Figure 2. Evolution of Mineralocorticoid, Glucocorticoid, Progesterone and Androgen Receptors.**
The corticoid receptor (CR) and PR are found in lampreys and hagfish, which are jawless fishes (cyclostomes) that evolved at the base of the vertebrate line. A separate MR and GR first appear cartilaginous fish (Chondrichthyes), such as sharks, rays and skates. The AR first appears in cartilaginous fishes.

We found that progesterone, as well as two related steroids, 19norProgesterone and spironolactone, are transcriptional activators of elephant shark MR [12], which we interpreted as indicating that progesterone was an ancestral activator of the MR and that progesterone activation of the MR was conserved in ray-finned fish [11-13, 15, 16, 52], lost in amphibians, alligators, rats and humans, and that progesterone activation of the MR evolved independently in chickens [12, 13].

**Evolution of progesterone binding to the mineralocorticoid receptor.**

An insight into the evolution of progesterone binding to the MR comes from an analysis of the sequences of steroid receptors in jawless vertebrates (Cyclostomes: lampreys, hagfish) and cartilaginous fish (Chondrichthyes: sharks, rays, skates, chimaeras) [1, 36, 53-58] (Figure 2). Lampreys and hagfish contain a corticoid receptor (CR), which is related to both the PR and MR [1, 36, 55, 58]. Sequence analysis reveals that the CR and PR evolved from a common ancestor [1, 36, 55, 58] in an ancestral jawless vertebrate that has descendants in modern lampreys and hagfish [57, 59] (Figure 2). The MR and its kin, the GR, first appear in cartilaginous fish [1, 36, 55] (Figure 2). The MR and the GR evolved through duplication and divergence of a CR [7, 28, 36, 60]. Sequence analysis indicates that the MR is closer to the CR than is the GR [1, 36, 55, 60].

The CR in jawless vertebrates is activated by progesterone [55]. Progesterone activation of the CR is conserved in elephant shark MR, which is activated by nM concentrations of progesterone [12] (Figure 3). Progesterone activation of elephant shark MR is conserved in ray-finned fish MRs [11-13, 15, 16, 52]. However, progesterone activation of the MR was lost in amphibians, alligators, rodents and humans [9, 13], in which aldosterone is the physiological mineralocorticoid [28, 61].

As mentioned earlier, aldosterone has not been found in either cartilaginous fish or ray-finned fish. Aldosterone first appears in lungfish [62-64], which are forerunners of terrestrial vertebrates. The evolution of aldosterone as the mineralocorticoid in amphibians, in which it regulates electrolyte homeostasis, has been proposed to be important in the conquest of land by terrestrial vertebrates [28, 65, 66]

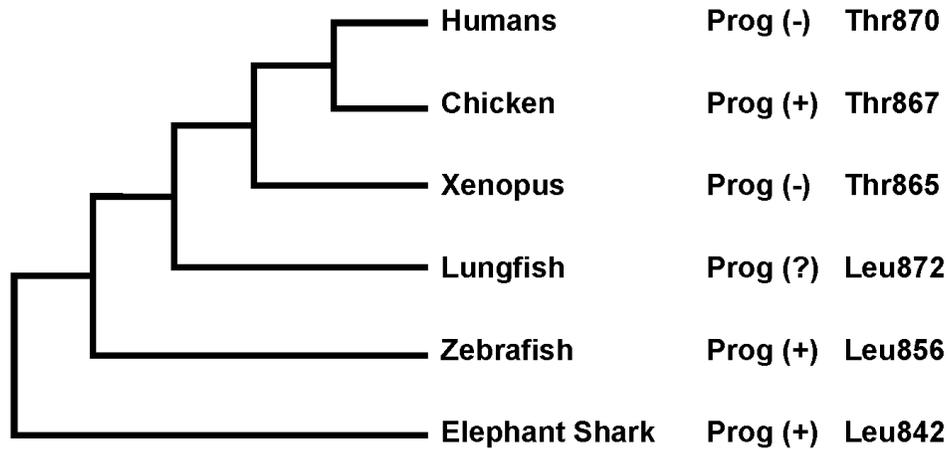

**Figure 3. Dependence of Activation by Progesterone of vertebrate MRs on the presence of either Leucine or Threonine in Helix 8.**
MRs with a key leucine in helix 8 are activated by progesterone. With the exception of chicken MR, progesterone does not activate MRs containing a threonine in helix 8.

**Mutant human MR Thr870Leu is activated by progesterone.**

For over a decade, the molecular mechanism for progesterone acting as an MR agonist in ray-finned fish and an MR antagonist in mammals was unknown [36]. Recently, Fuller et al. [52] solved this riddle through a series of experiments in which they studied progesterone activation of chimeras of human and zebrafish MR, in which segments of their LBDs were exchanged to form human and zebrafish chimeras that were investigated for activation by progesterone. These studies identified Leu856 in helix 8 in zebrafish MR as critical for progesterone activation of zebrafish MR (Figure 4). Replacement of Leu856 with threonine, which is the corresponding amino acid in helix 8 in human MR, yielded zebrafish Thr856 MR, which was not activated by progesterone. Importantly, replacement of Thr870 in human MR with leucine, yielded a mutant Leu870 MR that was activated by progesterone, as well as by aldosterone. These experiments confirmed the critical role of Thr870 in human MR in determining progesterone antagonist activity and of Leu856 in zebrafish MR in regulating progesterone agonist activity [52]. Interestingly, in mouse and rat MR, a serine corresponds to Thr870 of human MR. Substitution of serine for Leu856 in zebrafish MR results in the same loss of activation by spironolactone as found for substitution with threonine [52].

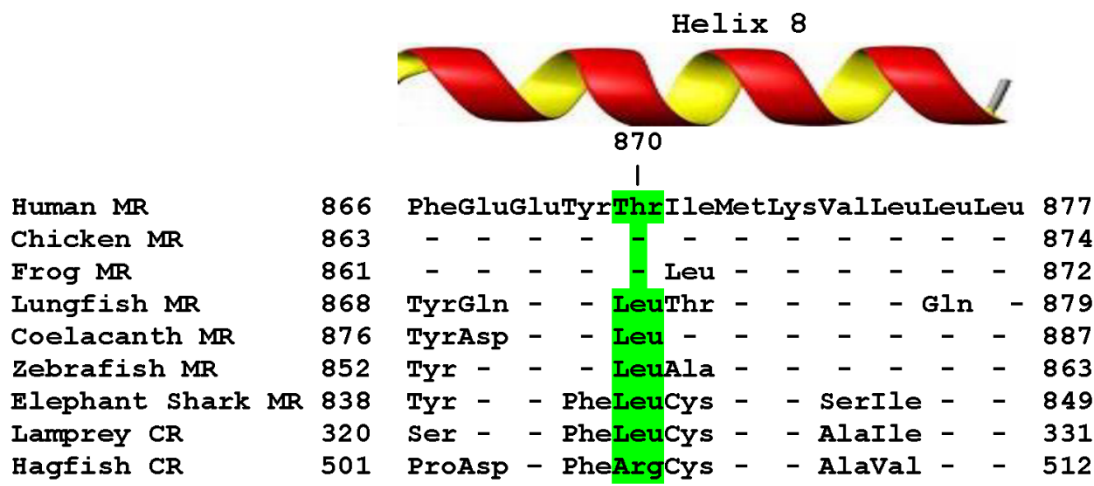

**Figure 4. Alignment of Helix 8 in the MR in Human, Chicken, Xenopus, Lungfish, Elephant Shark and Zebrafish and CR in Lamprey and Hagfish**.

Analysis of MR sequences in other vertebrates revealed that elephant shark MR and ray-finned fish MRs have a leucine corresponding the Leu856 in zebrafish MR, which explains their activation by progesterone [52]. Fuller et al. noted that Leu872 in lungfish MR (Figure 4) corresponds to zebrafish MR Leu856, suggesting that progesterone activates lungfish MR. Frog MR has Thr871 at this position, consistent with progesterone being an antagonist for frog MR [13]. Unexplained, at this time, is the basis for progesterone activation of chicken MR, which has Thr867 at the key position in helix 8. Thus, progesterone would be expected to inactivate chicken MR. It appears that another mechanism can lead to progesterone activation of chicken MR.

**Progesterone, a physiological ligand for cartilaginous fish and ray-finned fish MRs.**

Because aldosterone is not synthesized by ray-finned fish, other corticosteroids: cortisol, 11-deoxycorticosterone and corticosterone, have been proposed as physiological ligands for fish MRs [15, 67-71]. We proposed that progesterone also is a candidate activator of fish MR due to its low EC50 for MRs in zebrafish, gar and sturgeon [12, 16]. Moreover, synthesis of progesterone is simpler than 11-deoxycorticosterone (21-hydroxy-progesterone) and cortisol (Figure 5) providing a parsimonious model for the hypothesis that progesterone is a physiological ligand for ray-finned fish and cartilaginous fish MRs, although 11-deoxycorticosterone and other corticosteroids also may activate fish MRs.

We also note that progesterone lacks an 11β-hydroxyl, and thus progesterone is inert to 11β-hydroxysteroid dehydrogenase-type 2 (11β-HSD2). Fish contain 11β-HSD2 [28, 72, 73], which regulates access of cortisol to the MR [7, 27, 28, 35, 72-74]. Thus, like aldosterone and 11-deoxycorticosterone, progesterone can occupy the MR in the presence of 11β-HSD2, an enzyme that can act as gatekeeper to exclude cortisol and corticosterone from the MR.

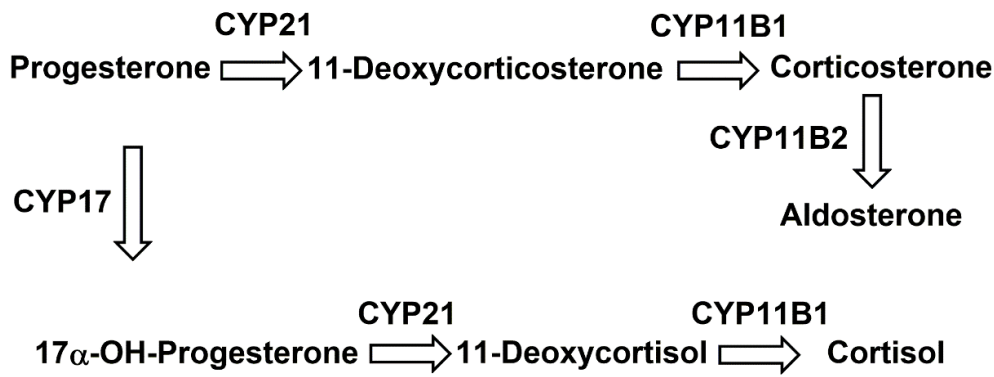

**Figure 5. Pathway for Synthesis of Adrenal Steroids from Progesterone.**
Cortisol is synthesized from progesterone in three steps.

**Unexplored functions of progesterone activation of cartilaginous fish and ray-finned fish MRs.**

Progesterone activation of fish MR indicates that progesterone has two functions in ray-finned fishes. One function, of course, is progesterone's classical activity as a reproductive hormone [75-77]. A second unexplored function is the physiology of progesterone activation of fish MR, which is expressed in diverse tissues in cichlids [78] and trout [15], including ovary a reproductive tissue in which progesterone also activates the PR. Moreover, analysis of gene transcription using RNA-Seq reveals that the MR in elephant shark is expressed in many tissues [12], including brain, heart, lung and liver, ovary and testis, as well as kidney and colon, the two "traditional" tissues for MR physiology. Little is known of the function progesterone activation of the MR in ovary, testis or other tissues in ray-finned fish or elephant shark.

Studies of corticosteroid activation of the MR in zebrafish [79-83] and medaka [70, 71, 84, 85] are beginning to elucidate the role of the MR in extra-renal tissues. We suggest parallel studies in zebrafish and medaka to investigate progesterone activation of the MR in heart, brain, lung, kidney, gill, liver, ovary, testis and other organs that contain an MR. For example, analysis of gene transcription using RNA-Seq of these organs in zebrafish exposed to progesterone, cortisol, 11-deoxycorticosterone and aldosterone should provide important insights into similarities and differences in gene expression due to these steroids.

The discovery by Fuller et al. [52] that zebrafish Thr856 MR is not activated by progesterone provides a neat protocol for studying the role of progesterone activation of zebrafish MR. CRISPR [86, 87] could selectively insert Thr856 into zebrafish MR in brain, heart, ovary, testis, kidney, gill, etc. to provide tissues in which the MR is not activated by progesterone, but can still be activated by circulating cortisol and deoxycorticosterone. This approach provides an opportunity to elucidate the role of progesterone in zebrafish, and it may uncover novel physiological functions for the MR in ray-finned fish, which are only beginning to be understood [71, 79-85] . Unexpectedly, although the MR is expressed in kidney and gill, the MR does not appear to regulate sodium uptake, the classical "mineralocorticoid function", in fish. Instead, the GR, activated by cortisol, regulates sodium uptake in zebrafish [88, 89], indicating that the MR has other functions in fish kidney and gill.

**Unexplored functions of progesterone in human MR**

The physiological consequences of progesterone acting as an antagonist of human MR are poorly understood [36, 49, 90-92]. An important advance in appreciating the physiological relevance of circulating levels of progesterone on human MR comes from Geller et al,'s [9] discovery of a human MR Ser810Leu mutation that is activated by progesterone. Geller et al [9, 93] found that a 15-year old male with MR Ser810Leu had high blood pressure (210/120 mmHg) due to sodium reabsorption mediated by progesterone activation of this mutant MR, indicating that physiological levels of progesterone in males are sufficient to activate this mutant MR. As expected, females with Leu810 MR also have high blood pressure. Considering the nM affinity of progesterone for human MR and progesterone's inertness to 11β-HSD2, progesterone would be expected to occupy wild-type MR in tissues expressing 11β-HSD2 [28, 29, 35, 38, 41, 94, 95].

Cortisol has been proposed to activate human MR in tissues that lack 11β-HSD2. A role for progesterone in competing with cortisol for binding to the MR in these tissues needs to be investigated.

**Progesterone activation of chicken MR.**

Progesterone activation of chicken MR was unexpected [12, 13] because chicken MR, Thr667 corresponds to Thr670 in human MR, which should mean that chicken MR is not activated by progesterone. The mutation(s) that confer progesterone activation of chicken MR are unknown. The protocol used by Fuller et al. [52] to decipher the basis for progesterone activation of zebrafish MR should work to decipher progesterone activation of chicken MR.

The use of RNA-Seq analysis proposed for studying progesterone activation of zebrafish MR also can be used to elucidate the function of progesterone activation of chicken MR in brain, heart, lung, thyroid, liver, kidney, spleen, stomach, pancreas, and pituitary [96]. RNA-Seq analysis of these tissues from chickens exposed to progesterone and aldosterone should provide insights into similarities and differences in gene expression due to these steroids.

**Funding**: This work was supported in part by Grants-in-Aid for Scientific Research 19K0673409 (YK) from the Ministry of Education, Culture, Sports, Science and Technology of Japan. M.E.B. was supported by Research fund #3096.
**Author contributions**: M.E.B. and Y.K. wrote the paper.
**Competing interests**: We have no competing interests.